\documentclass[fleqn,usenatbib]{mnras}

\usepackage{newtxtext,newtxmath}
\usepackage{txfonts}
\usepackage{ulem}

\usepackage[T1]{fontenc}

\DeclareRobustCommand{\VAN}[3]{#2}
\let\VANthebibliography\thebibliography
\def\thebibliography{\DeclareRobustCommand{\VAN}[3]{##3}\VANthebibliography}

\usepackage{multirow}
\newcommand\Cal{CAL\,83}
\newcommand\NH{$N_{\text{H}}$}

\usepackage{xcolor}

\usepackage{graphicx}	
\usepackage{amsmath}	
\usepackage{amssymb}	
\usepackage{import}

\title[Revisiting \Cal]{Revisiting multiwavelength data on the supersoft X-ray source \Cal}

\author[Stecchini et al.]{
P. E. Stecchini$^{1,2}$\thanks{E-mail: paulo.stecchini@usp.br / paulo.stecchini@inpe.br},
M. P. Diaz$^{1}$,
F. D'Amico$^{2}$,
and F. Jablonski$^{2}$\vspace{0.4cm}\\
\parbox{\textwidth}{$^{1}$IAG, Universidade de São Paulo, Rua do Matão 1226, 05508–900,  São Paulo--SP, Brazil\\
$^{2}$Divisão de Astrofísica, Coordenação de Engenharia, Tecnologia e Ciências Espaciais, Instituto Nacional de Pesquisas Espaciais, Av. dos Astronautas 1758, 12227-010, S.J. Campos--SP, Brazil\\
}}

\date{Accepted XXX. Received YYY; in original form ZZZ}

\pubyear{2021}

\begin{document}
\label{firstpage}
\pagerange{\pageref{firstpage}--\pageref{lastpage}}
\maketitle

\begin{abstract}
In this study we revisit public data on the supersoft X-ray source \Cal\space in the Large Magellanic Cloud. A significant part of our analysis is focused on \textit{XMM-Newton} X-ray observations, in which updated data reduction procedures and quality assessment were applied. We report on the capability of publicly available hot atmosphere models in describing the source's soft X-ray spectrum. By gathering historical flux measurements in multiple wavelengths and comparing them with the fluxes derived from the X-ray analysis, we find that a $\sim$\,360\,kK phenomenological blackbody model describes the spectral energy distribution of \Cal\space fairly well. We also retrieve data from the \textit{XMM-Newton} UV/optical camera, which is co-alligned with the X-ray instruments and provides strictly simultaneous measurements. These observations demonstrate that the X-ray emission is definitely anti-correlated with emission at longer wavelengths in a time-scale of days to weeks. A closer look at simultaneous X-ray and UV count rates in single light curves reveals that the anti-correlated behaviour is actually present in time scales as short as minutes, suggesting that the origin of variable emission in the system is not unique. 
\end{abstract}




\section{Introduction}
\label{sect:01}

Objects today known as supersoft X-ray sources (SSS) were first observed by the \textit{Einstein Observatory} during a Large Magellanic Cloud (LMC) soft X-ray survey conducted by \cite{1981ApJ...248..925L}. \Cal\space and CAL\,87, with their unusual soft emission amongst over 90 sources catalogued by the Columbia Astrophysics Laboratory team (hence the source names), inaugurated the class of SSS, becoming their prototypes. A subsequent survey in the LMC with \textit{ROSAT} confirmed their very soft X-ray emission and added RX\,J0527.8-6954 \citep{1991A&A...246L..17G, 1991Natur.349..579T} to the list; to date, hundreds of objects have been observed to exhibit similar behaviour and are candidates or confirmed members of the class. The list now contains sources located not only in the LMC but also in the Milky Way, Small Magellanic Cloud, M31, M33 and other galaxies \citep[e.g.][]{1996IAUS..165..425K,  1998PASP..110..276S, 2000NewA....5..137G, 2003AN....324...85P, 2003ApJ...592..884D, 2010ApJ...717..739O, 2010AN....331..193H, 2012A&A...537A..76S, 2021A&A...646A..85G}.

Common aspects present in most supersoft sources include -- but are not limited to -- very soft X-ray spectra (with weak or no emission beyond $\sim$\,1\,keV), effective blackbody temperatures of no more than $\sim$\,100\,eV and bolometric luminosities of $\sim$\,10$^{36-38}$\,erg\,s$^{-1}$ \citep[e.g][]{1997ARA&A..35...69K}. The precise origin of these features and the nature of these objects were a matter of debate in the following years after the prototypes' identification \citep[e.g.][]{1990ApJ...350..288C, 1991A&A...246L..17G}, until \cite{1992A&A...262...97V} pointed out that the observed emission was consistent with the scenario of a steady hydrogen burning massive ($\gtrsim$\,1\,M$_{\odot}$) white dwarf (WD) that accretes matter with rate of the order of 10$^{-7}$\,M$_{\odot}$\,year$^{-1}$ from its likely even more massive ($\sim$\,1--3\,M$_{\odot}$) companion. Modelling the formation and evolution of WDs in this scenario have successfully reproduced typical observational parameters \citep[e.g.][]{1994ApJ...426..692R, 2004ApJ...601.1058I}.
Similar X-ray spectral shapes have been observed in one of the evolutionary stages of novae outbursts (during the so-called supersoft phase), which corroborates the association of the SSS phenomenology with WDs in binary systems \citep[e.g.][]{2006ApJS..167...59H, 2010AJ....140.1860D, 2018ApJ...862..164O}. In supersoft X-ray sources, however, the high mass transfer rates and steady burning of hydrogen prevent degenerate thermonuclear runaways from occurring \citep[e.g.][]{2013ApJ...777..136W}, making it possible for these objects to achieve masses close to the Chandrasekhar limit ($\approx$\,1.4\,M$_{\odot}$) and thus become important candidates to type Ia Supernova progenitors \citep[e.g.][]{2003ApJ...590..445H, 2007ApJ...663.1269N, 2010AN....331..218P}.

The optical flux observed in SSS is mostly credited to an accretion disk around the WD. The inferred luminosities for some systems, however, seem to exceed the expected for regular cataclysmic variable disks and are attributed to X-ray photons that originate in the hot WD's atmosphere and are reprocessed in the accretion disk -- rather than produced by viscous accretion processes \citep{1996LNP...472...65P}. It has been reported for a couple of fairly well monitored supersoft sources [e.g. RX\,J0513.9--6951 \citep[e.g][]{1993A&A...278L..39P, 2005MNRAS.364..462M} and \Cal\space\citep[e.g][]{2002A&A...387..944G, 2013MNRAS.432.2886R}] that the rise of optical -- and also ultraviolet (UV) -- fluxes occur almost concomitantly with the decline of X-ray emission. Interpretations for the mechanisms responsible for this anti-correlation vary in the literature, but most are based on the expansion/contraction model proposed by \cite{1996ApJ...470.1065S}. \citeauthor{1996ApJ...470.1065S} argue that X-ray faint and optical bright states begin after the mass accretion rate of the system increases, causing the WD's photospheric radius to expand which, in turn, also lowers the effective temperature of the burning shell and shifts the emission peak from X-rays to longer wavelengths (i.e extreme UV or UV). Inversely, a decrease in the accretion rate contracts the photosphere and enhances the X-ray emission.

\Cal, subject of this article, is one of the most studied supersoft X-ray sources. The very massive nature of the WD (possibly up to 1.3\,M$_{\odot}$) -- that is currently a consensus -- was firstly proposed by \cite{1997MNRAS.286..483A} and \cite{1998A&A...331..328K} after modelling the short timescale in which the source went from an X-ray bright to an off state, and further corroborated by a high resolution X-ray spectroscopic analysis performed by \cite{2005ApJ...619..517L}. The nature of the donor star in \Cal, none the less, is still a debated aspect of the system. 
Early optical spectroscopy and photometry revealed a periodic modulation of about 1~day \citep{1987ApJ...321..745C,1988MNRAS.233...51S}, value later confirmed, refined ($\sim$\,1.0475\,d) and well established as the orbital period of the binary system \citep{2004AJ....127..469S,2013MNRAS.432.2886R}. The absence of eclipses and the semi-amplitude of the radial velocity ($\sim$\,25--40\,km\,s$^{-1}$) curves in the He\,\textsc{ii} (4686\,$\AA$) emission line throughout an entire orbital phase \citep[e.g.][]{1987ApJ...321..745C,2004AJ....127..469S} suggest that the system has a low orbital inclination ($<$\,30$^{\circ}$). With these two constraints, and acknowledging that the WD is likely massive, the mass of the secondary could be as low as $\sim$\,0.5\,M$_{\odot}$ \citep[e.g.][]{2014MNRAS.437.2948O},  a value significantly smaller than that required by \cite{1992A&A...262...97V} model to allow steady thermonuclear burning under unstable mass transfer. 

From X-ray temporal studies, two modulations have been identified: a 38.4\,min variability in \textit{Chandra}-LETG data \citep{2006AJ....131..600S} and, from \textit{XMM-Newton} \citep{2014MNRAS.437.2948O} and \textit{NICER} \citep{2022ApJ...932...45O} observations, a (sometimes) prominent -- but not very stable -- signal of $\sim$\,67\,s. The latter has been associated to the WD's spin period. The X-ray emission of \Cal\space is very soft -- even amongst SSS --, with only few counts above $\sim$\,0.5\,keV.  
Both blackbody or WD atmosphere models have been shown to satisfactorily describe the source's X-ray continuum (i.e. low resolution spectra), with best-fitting effective temperatures ranging roughly from 20 to 50\,eV \citep[e.g][]{1991A&A...246L..17G, 1998A&A...332..199P, 2013MNRAS.432.2886R}. A so-called X-ray off state of \Cal\space was first noticed by \cite{1997ASPC..121..730K} from a \textit{ROSAT} observation; since then, the source has been reported to be undetectable in X-rays on seven other occasions \citep[see Table~2 of][]{2013MNRAS.432.2886R}. The MACHO project long-term optical monitoring of \Cal\space allowed \cite{2002A&A...387..944G} and \cite{2013MNRAS.432.2886R} to conclude that these X-ray off periods all happened while the source was in its optical bright state and vice-versa (bright in X-rays while faint in optical).

In this study we revisit most available \textit{XMM-Newton} \citep{2001A&A...365L...1J} observations of \Cal\space to, through different approaches, explore its X-ray spectrum, its temporal variability and possible correlations with emission at other wavelengths.
In Section\,\ref{sect:02}, we detail the criteria, reduction tasks and procedures adopted to process the X-ray data; measurements in other wavelengths used in our analysis will be explained as they are mentioned. In Section\,\ref{sect:03}, we describe the analysis tools, models and approaches used, as well as present and briefly comment on the results, whose relevance for \Cal\space are more acutely discussed in Section\,\ref{sect:04}. The main outcome of our study is presented in Section\,\ref{sect:05}.

\section{Data reduction}
\label{sect:02}

The region of \Cal\space was observed by the instruments on-board the \textit{XMM-Newton} satellite on 22 occasions from 2007 to 2009. We retrieved data from these observations for the three X-ray cameras, EPIC-pn \citep{2001A&A...365L..18S} and EPIC-MOS 1, 2 \citep{2001A&A...365L..27T}. However, as we are dealing with a very soft X-ray source and thus spectra with the lowest energy possible are desired, we chose to use only data from the pn detector due to its higher sensitivity at lower energies compared to that of the MOS detectors. Also, following the recommendations of the latest calibration status document\footnote{\url{https://xmmweb.esac.esa.int/docs/documents/CAL-TN-0018.pdf}}, pn data analysis was restricted to energies $\geq$\,0.2\,keV. 
Data reduction procedures to be described were all conducted with the \textit{XMM-Newton} Science Analysis System (SAS, v.\,19.1.0).

After downloading the Observation Data Files (ODF) from the \textit{XMM-Newton} Science Archive\footnote{\url{http://nxsa.esac.esa.int/}}, we ran the preparatory tasks \texttt{cifbuild} and \texttt{odfingest}. The former produces the Calibrated Index File (CIF), which associates the Current Calibration File\footnote{\url{https://www.cosmos.esa.int/web/xmm-newton/current-calibration-files}} (CCF) to each particular observation; the latter extends the ODF summary, now to include these instrumental and calibration informations. To reprocess the ODFs and generate calibrated (yet not filtered) event files, the task \texttt{epproc} was used. Parameter \texttt{runepreject=yes} was set in order to mitigate detector noise at low energies \citep[e.g.][]{2004SPIE.5488...61D}.   
Possible flaring particle background was filtered out from the event files by applying the standard selection criteria\footnote{\url{https://www.cosmos.esa.int/web/xmm-newton/sas-thread-epic-filterbackground}}. The output are cleaned event files ready for science products (spectra, light curve) extraction.  

The source photons from each observation were extracted from circular regions of 32 arcsec radius centred on \Cal\space centroid position. This radius encompasses about 85--90\% of the telescope's point spread function encircled energy\footnote{\url{https://xmm-tools.cosmos.esa.int/external/xmm_user_support/documentation/uhb/onaxisxraypsf.html}}. The background regions were chosen with the support of the SAS task \texttt{ebkgreg}\footnote{\url{https://xmm-tools.cosmos.esa.int/external/sas/current/doc/ebkgreg/}}. This task indicates the optimal background extraction region based solely on the detector's geometry; hence, in some occasions, the suggested regions had to be slightly shrunk to avoid the inclusion of photons either from the source itself or from other objects in the field-of-view. The resulting background regions were all circular, adjacent to the source and with radii varying from 40 to 60 arcsec. For four observations (ObsIDs 0500860701, 0500860801, 0506530301, 0506530401) there were no statistically significant counts within the source region. 
Basic information on the 22 observations is listed in Table\,\ref{tab:01}.

\begin{table}

\caption{Summary of the observations.}
\label{tab:01}

\begin{tabular}{ccccc}

\hline
\multicolumn{1}{c}{Observation} & \multicolumn{2}{c}{Date}               & \multicolumn{1}{c}{Exposure$^a$} & \multicolumn{1}{c}{Count rate$^b$} \\ 
\multicolumn{1}{c}{(ObsID)}     & \multicolumn{1}{c}{(MJD)$^c$} & (dd/mm/yyyy) & \multicolumn{1}{c}{(s)}      & \multicolumn{1}{c}{(cps)}      \\ \hline
0500860201                      & 54233.92                  & 13/05/2007 & 11471                        & 3.80                           \\ 
0500860301                      & 54287.99                  & 06/07/2007 & 10470                        & 3.56                           \\
0500860401                      & 54333.64                  & 21/08/2007 & 7471                         & 2.74                           \\
0500860501                      & 54379.05                  & 06/10/2007 & 12506                        & 2.63                           \\
0500860601                      & 54428.93                  & 24/11/2007 & 20030                        & 3.39                           \\
0500860701                      & 54481.56                  & 16/01/2008 & 10471                        & -                              \\
0500860801                      & 54535.43                  & 10/03/2008 & 6470                         & -                              \\
0506530201                      & 54545.03                  & 20/03/2008 & 7271                         & 0.54                           \\
0506530301                      & 54559.78                  & 03/04/2008 & 14472                        & -                              \\
0506530401                      & 54567.26                  & 11/04/2008 & 5471                         & -                              \\
0506530501                      & 54572.59                  & 16/04/2008 & 4594                         & 1.11                           \\
0506530601                      & 54573.57                  & 17/04/2008 & 10765                        & 0.90                           \\
0506530801                      & 54575.28                  & 19/04/2008 & 5471                         & 0.33                           \\
0506530901                      & 54576.94                  & 20/04/2008 & 11170                        & 0.30                           \\
0500860901                      & 54577.09                  & 21/04/2008 & 12269                        & 0.18                           \\
0506531001                      & 54577.79                  & 21/04/2008 & 8668                         & 0.75                           \\
0506531201                      & 54579.48                  & 23/04/2008 & 6971                         & 0.20                           \\
0506531301                      & 54581.35                  & 25/04/2008 & 9171                         & 0.38                           \\
0506531401                      & 54585.03                  & 29/04/2008 & 13669                        & 0.21                           \\
0506531501                      & 54690.62                  & 12/08/2008 & 6471                         & 4.82                           \\
0506531601                      & 54726.47                  & 17/09/2008 & 6371                         & 0.12                           \\
0506531701                      & 54981.34                  & 30/05/2009 & 45664                        & 4.77                           \\ \hline
\end{tabular}
{\begin{flushleft}\small{} $^a$The sum of good-time-invervals (from keyword \texttt{ONTIME}).

$^b$Total counts in source region divided by the exposure time.

$^c$Modified Julian Date.
\end{flushleft}}
\end{table}

Before proceeding to the spectra extraction, we investigated if the observations were affected by pile-up. We first followed the procedure described in \cite{2015A&A...581A.104J} to check whether the count rate of any observation exceeded the limits in which the pile-up effects may become worrisome. Even for the observation with the highest mean rate (ObsID 0506531501; 4.88\,cps in source region), we calculated spectral distortion and flux loss of less than $\sim$\,0.2\% and $\sim$\,0.8\%, respectively. Additionally -- and as advised in the SAS Threads --, we also ran, for all observations, the pile-up diagnosis tool \texttt{epatplot}\footnote{\url{https://www.cosmos.esa.int/web/xmm-newton/sas-thread-epatplot}}, which provides the observed-to-model ratios for single and double pattern events. The energy range of evaluation was extended from 0.5 (default) down to 0.2\,keV (minimum energy intended for analysis) and parameter \texttt{withbackgroundset=Y} was passed, as low energy background may also cause deviations in the pattern fractions. Overall outcome was that both single and double events observed are in good agreement with the expected (i.e. pattern fractions follow the model curves), doubles being slightly \textit{deficient} in some cases.
An \textit{excess} of double over single events would be indicative of a piled-up observation, whilst the opposite (singles over doubles) could be related to effects of X-ray/optical loading\footnote{\url{https://xmmweb.esac.esa.int/docs/documents/CAL-TN-0050-1-1.pdf}}. It is worth noting, though, that for very soft sources like \Cal, the models used to estimate the number of singles and doubles might not be accurate and the interpretation of \texttt{epatplot} output may not be that straightforward\footnote{Private communication: \textit{XMM-Newton} HelpDesk (\url{https://xmmweb.esac.esa.int/xmmhelp/})}. To leave no doubt about the presence of either pile-up or X-ray loading, we reprocessed (\texttt{epproc}) the raw data with the respective correction tasks \texttt{pileuptempfile=yes} and \texttt{runepxrlcorr=yes} included. A comparison of the final spectra extracted with and without these two tasks (or any of them) reveals that deviations are within statistical uncertainties; we checked that these do not affect the results to be presented in the following sections. 

\begin{figure}
	\includegraphics[width=\columnwidth]{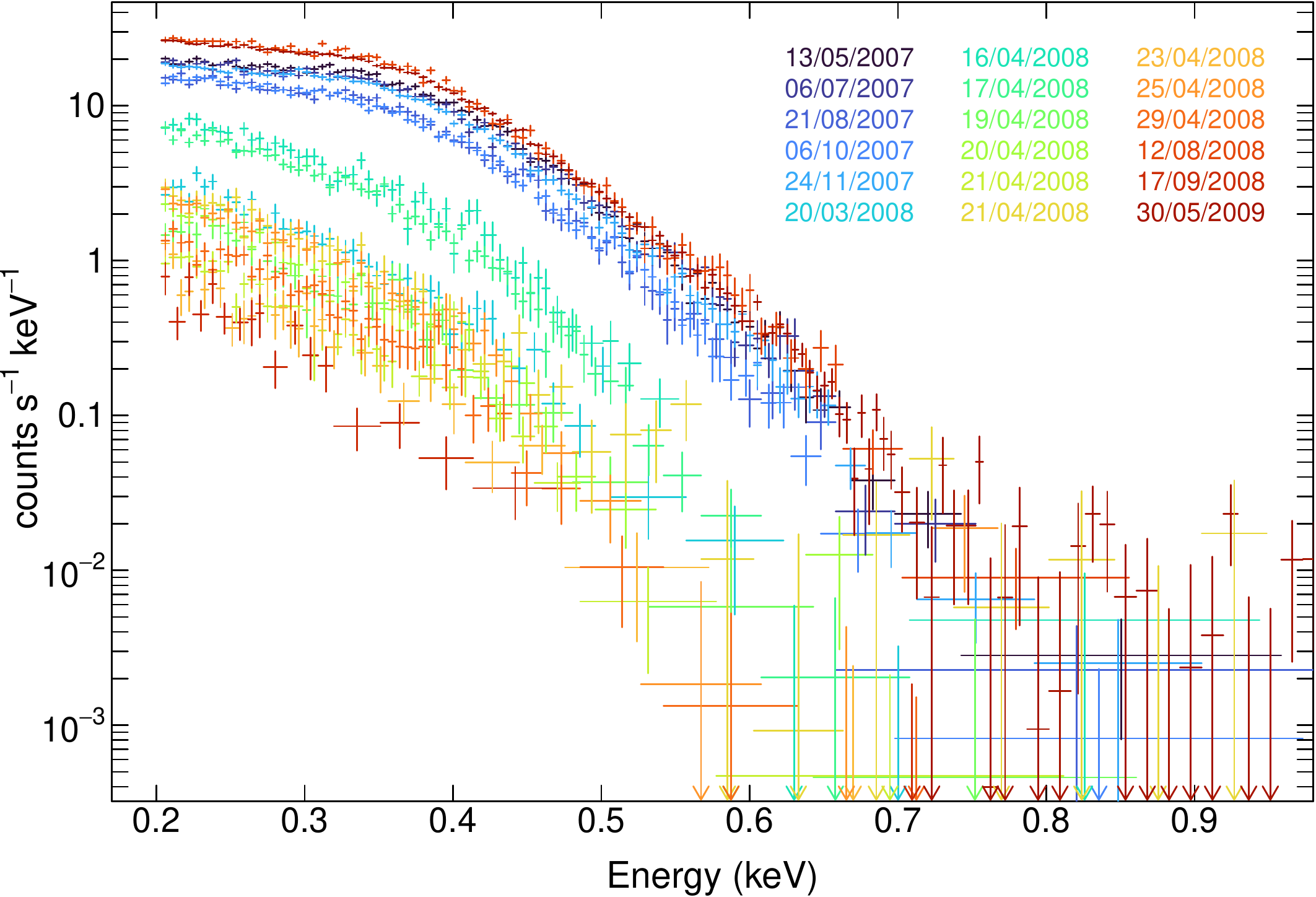}
    \caption{\textit{XMM-Newton} EPIC-pn spectra of \Cal\space for the 18 occasions in which the source was detectable. The corresponding observation dates are presented in the format dd/mm/yyyy. Down arrows represent upper limits.}
    \label{fig:01}
\end{figure}

Finally, we extracted the source and background spectra from the region files created; for maximising energy calibration and resolution, we chose to use only single events by setting \texttt{PATTERN==0}. 
Standard tasks \texttt{rmfgen} and \texttt{arfgen} were run to compute the redistribution matrix file (RMF) and the ancillary response file (ARF), respectively. Each spectrum was binned to have at least 10 counts per bin prior to analysis. The final spectra, for the 18 observations with enough counts within the source region, are shown in Figure\,\ref{fig:01}. Colours follow a sequence according to the observation dates, which are also displayed.

Even though we do not focus on timing analysis, to allow us to discuss on \Cal\space variability, we also extracted light curves (with the same source and background regions aforementioned) for different energy bands. For that, we used the SAS task \texttt{epiclccorr}\footnote{\url{https://heasarc.gsfc.nasa.gov/docs/xmm/sas/help/epiclccorr/}}, which performs a series of corrections to minimise effects that may impact the detection efficiency before producing a background-subtracted light curve. Time bins of 5 seconds were used.

Detailed description of aims and usage of all procedures and tasks mentioned (and not explicitly referenced) can be found in the SAS Threads\footnote{\url{https://www.cosmos.esa.int/web/xmm-newton/sas-threads}} or in the \textit{XMM-Newton} documentation\footnote{\url{https://xmm-tools.cosmos.esa.int/external/sas/current/doc/}}.

\section{Analyses \& Results}
\label{sect:03}

The spectral analysis was carried out with \textsc{XSPEC} \citep[][v.\,12.12.0]{1996ASPC..101...17A}. Energies below 0.2 and above 1.0\,keV were ignored for the model fitting. For all fits, two photoelectric absorption components were included: a Galactic column density (model \texttt{tbabs} from \textsc{XSPEC}) fixed at 6.5\,$\times$\,10$^{20}$\,cm$^{-2}$ \citep[as derived by][]{1998A&A...333..163G} with solar abundances and, to account for any intrinsic absorption of the LMC or the source, an additional component (model \texttt{vphabs}) with column density as a free parameter and abundance values (relative to solar) taken from Table~7 of \citet{2002A&A...396...53R}. Solar abundance was set according to \citet{2000ApJ...542..914W} and X-ray atomic cross-sections according to \citet{1996ApJ...465..487V}.

In addition to applying a phenomenological blackbody model to the 18 spectra (see Figure\,\ref{fig:01}), we also tested a few publicly available\footnote{\url{http://astro.uni-tuebingen.de/~rauch/TMAF/TMAF.html}} NLTE (non-local thermal equilibrium) stellar atmosphere models \citep{1999JCoAM.109...65W, 2003A&A...403..709R, 2003ASPC..288..103R, 2010ApJ...717..363R}. Some of them are available as tables that are suitable for analysis within the \textsc{XSPEC} environment; in general, these models -- or pre-calculated grids -- differ from each other by the abundance of elements, by their temperature and by the effective gravity (expressed as log\,$g$, in cgs units). Once the specific abundance and gravity are chosen, the fitting to the data is parameterised by the effective temperature. 

\begin{figure}
	\includegraphics[width=\columnwidth]{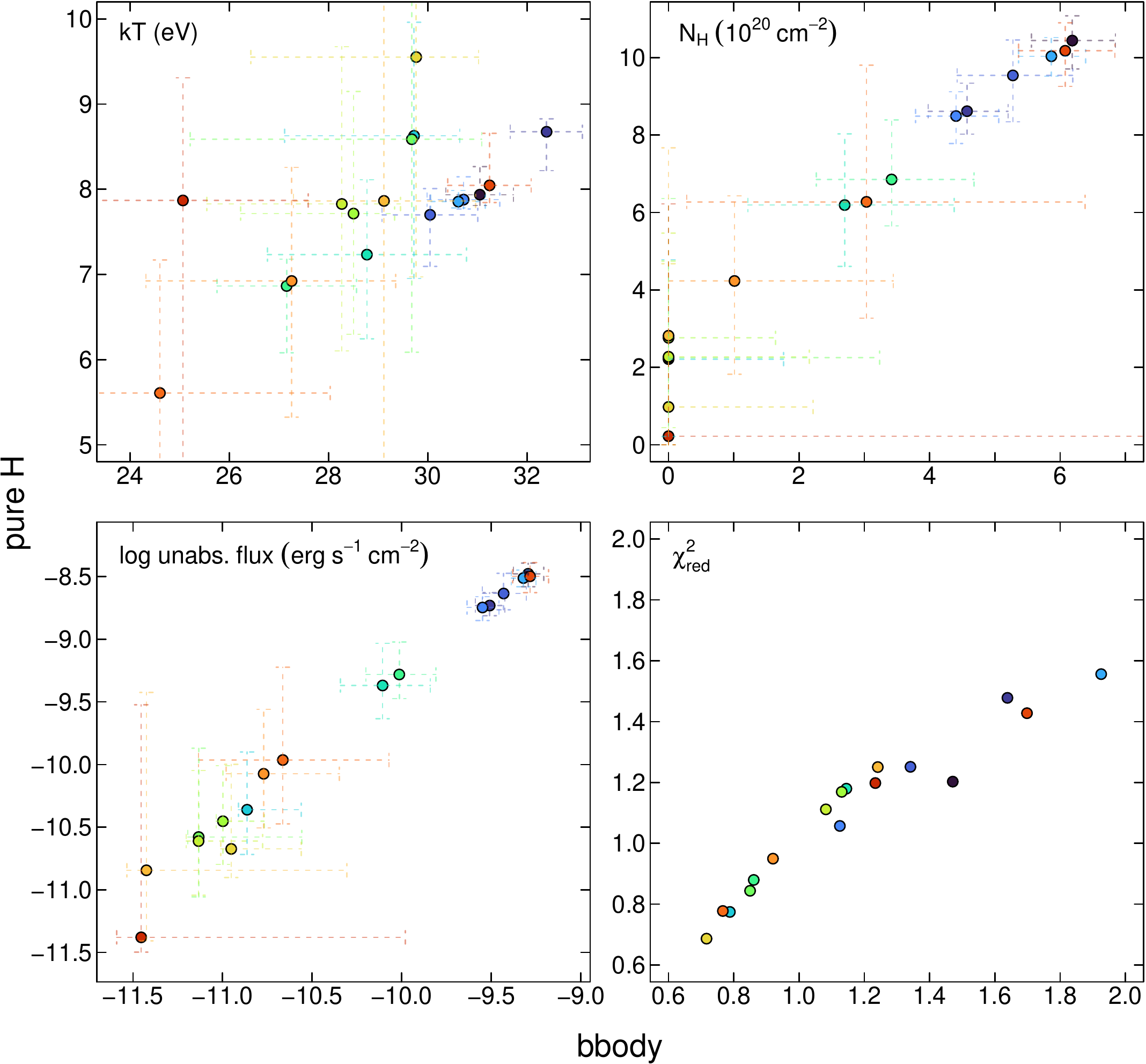}
    \caption{Best-fitting parameter values (and their associated 90\% error) provided by the blackbody (\texttt{bbody}) model (horizontal axis of all panels) and by the pure hydrogen  atmosphere (\texttt{pure\,H}) model (vertical axis of all panels). Each point in any panel is an observation, colour coded as in Figure\,\ref{fig:01}. \textit{Upper left}: effective temperature. \textit{Upper right}: equivalent hydrogen column with LMC abundances. \textit{Bottom left}: unabsorbed flux from 0.2 to 1\,keV. \textit{Bottom right}: fit quality (reduced chi-squared).}
    \label{fig:02}
\end{figure}

\begin{figure}
	\includegraphics[width=\columnwidth]{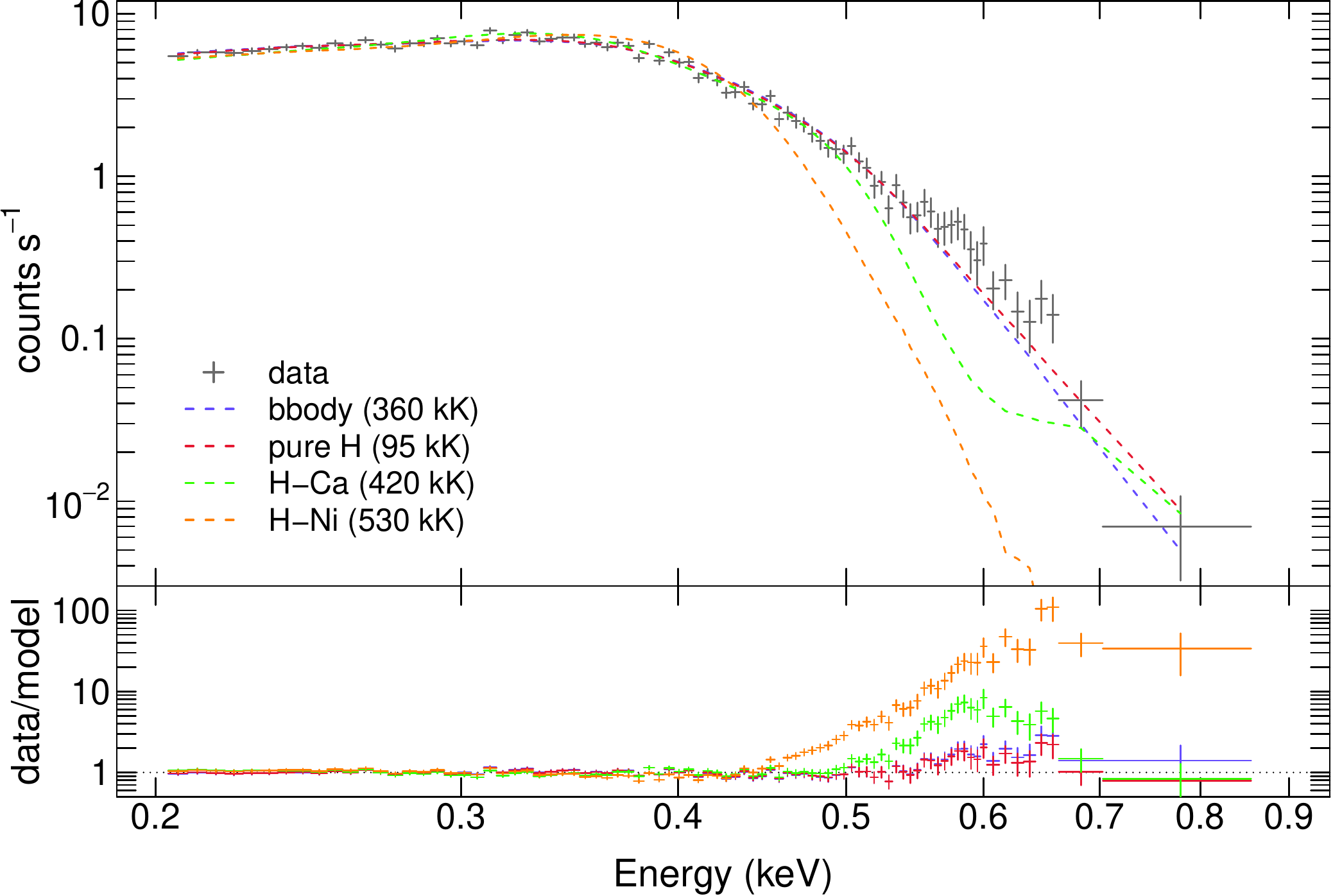}
    \caption{Different models fitted to data of ObsID 0506531501 (upper panel) and the corresponding residuals (bottom panel). Best-fitting temperatures for each model are shown. }
    \label{fig:03}
\end{figure}

All 18 spectra but one (ObsID 0506531701) could be satisfactorily fit (reduced chi-square $\chi^2_{\text{red}}$\,$\equiv$\,$\chi^2$/$\nu$\,$\leq$\,2, where $\chi^2$ is the chi-square statistics and $\nu$ is the number of degrees of freedom) by the blackbody model (\texttt{bbody}). Amongst the NLTE stellar atmosphere models, only the one with a pure hydrogen atmosphere (henceforth "\texttt{pure\,H}") and log\,$g$\,=\,7 achieved a similar fit quality. Figure\,\ref{fig:02} shows, in a compact way, the best-fitting parameters from these two models, for the 17 spectra in which $\chi^2_{\text{red}}$ was below 2. Each panel of the figure displays the output values (\texttt{bbody} in horizontal and \texttt{pure\,H} in vertical axes) for one specific parameter. Although the results from both models follow a consistent trend with respect to each other, it is important to notice that, except for the last panel, the axes scales are not the same. For instance, the effective temperatures are very different, with median values of 30\,$\pm$\,2\,eV ($\sim$\,320--370\,kK) for the \texttt{bbody} and  8\,$\pm$\,1\,eV ($\sim$\,80--105\,kK) for the \texttt{pure\,H} model. Further, the latter also requires larger values for the 
LMC column density (\NH) and provides, consequently, larger unabsorbed fluxes. At last, the fit quality (bottom right panel), when measured by the distance to the 1:1 relationship, is slightly better overall for the pure hydrogen atmosphere model. Both models fitted to data of ObsID 0506531501 (the one with the highest count rate) and the respective residuals are shown in Figure\,\ref{fig:03}, with best-fitting temperatures indicated. 

\begin{figure}
	\includegraphics[width=\columnwidth]{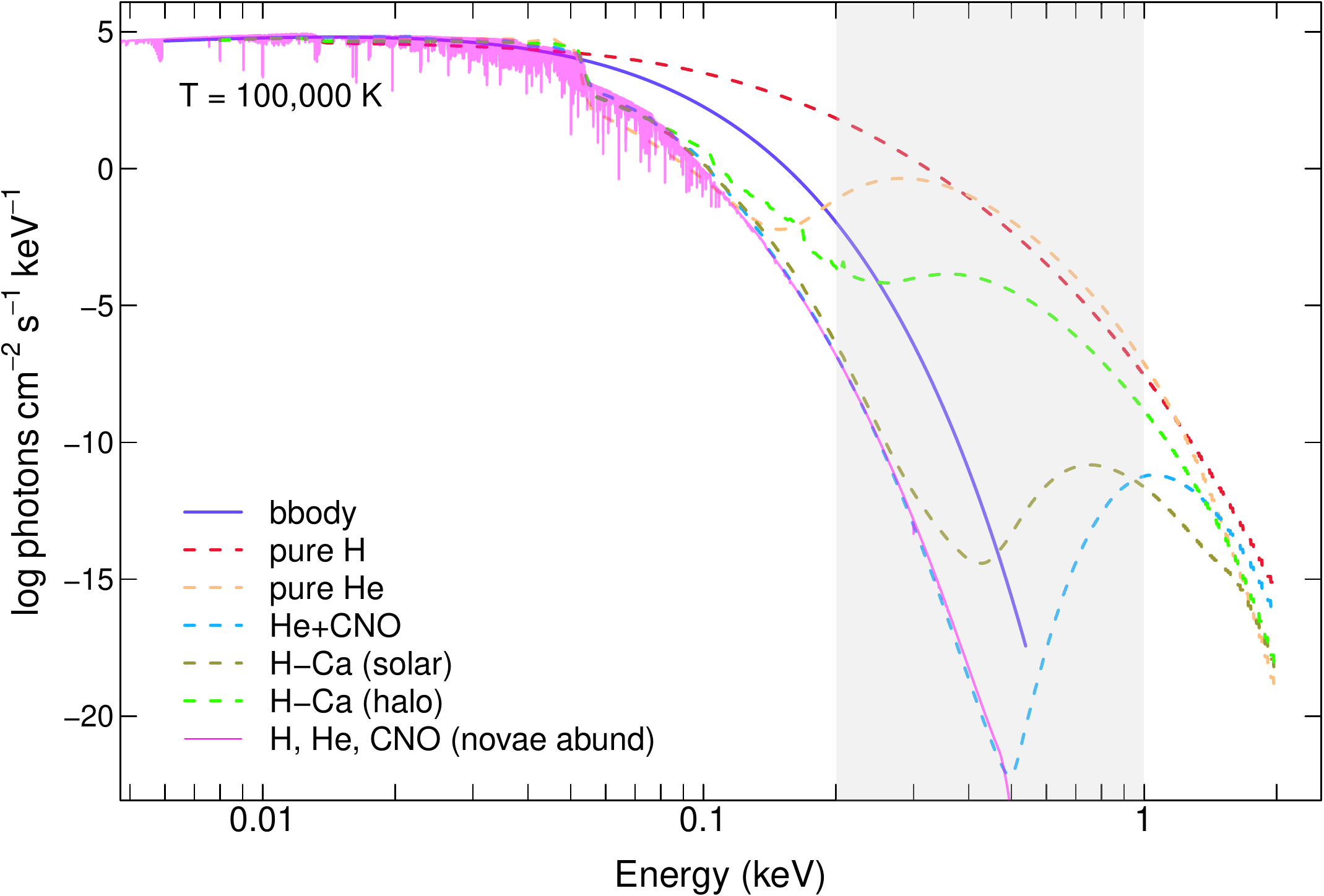}
    \caption{Photon flux as a function of energy of theoretical models calculated for T\,=\,100,000\,K, log\,$g$\,=\,7 (when applicable) and no interstellar absorption. They have been adjusted to match the \texttt{bbody} flux at 0.01\,keV. Energy range of analysis (0.2--1.0\,keV) is highlighted in grey.}
    \label{fig:04}
\end{figure}

Also plotted in Figure\,\ref{fig:03} are the fit to data by two other NLTE models: "\texttt{H--Ca}", that includes elements from H to Ca with abundance ratios as the Galactic halo and "\texttt{H--Ni}", with elements up to Ni and a particular metal enhanced abundance\footnote{Model series "003", \url{http://astro.uni-tuebingen.de/~rauch/TMAF/flux_HHeCNONeMgSiS_gen.html}} based on nova V4743\,Sgr \citep{2010arXiv1011.3628R, 2010ApJ...717..363R}. Even though aware that the \texttt{H--Ca} model was calculated with only approximate formulae to account for Stark broadening and is thus not suitable for \textit{precise} spectral analysis, we reckon that this should not be an issue given the low energy resolution of the spectra we are analysing. The \texttt{H--Ni} pre-calculated grid is only available for log\,$g$\,=\,9, so we use this effective gravity for the \texttt{H--Ca} as well. 
While a reasonable agreement between data and model is obtained from \texttt{bbody} or \texttt{pure\,H}, neither of the two other NLTE models are able to fit the data, as evidenced by the residuals, plotted in logarithm scale for clarity. Particularly, they notably underestimate the observed spectrum for energies beyond $\sim$\,0.45--0.5\,keV. It can also be noticed that they demand a much larger effective temperature -- specially when compared to the \texttt{pure\,H} model --, likely due to the absorption caused by elements heavier than helium.   
Merely to illustrate the models' sensitivity to the choice of elements and their abundances, we exhibit in Figure\,\ref{fig:04} a few theoretical models computed for T\,=\,100\,kK (temperature most commonly available amongst the models). Besides previously mentioned models \texttt{bbody}, \texttt{pure\,H} and \texttt{H--Ca} (with halo abundances), we also show some other NLTE (from those publicly available) computed for different abundances: "\texttt{pure\,He}", a pure helium atmosphere model; "\texttt{H--Ca}", this time with solar abundances; and "\texttt{He+CNO}", a model with abundance ratios seen in some PG\,1159 type objects \citep[He:C:N:O = 33:50:2:15, e.g.][]{2006PASP..118..183W}. In addition, we plot a fine wavelength sampled model we calculated with the \textsc{TLUSTY} code for a NLTE line-blanketed atmosphere, \citep{1995ApJ...439..875H, 2011ascl.soft09022H} aiming to show the impact of metal abundances on extreme/far UV blanketing at this temperature. This model, labelled as "\texttt{H, He, CNO}", was calculated with abundance ratios provided by a compilation of median abundances measured in 25 Galactic novae \citep[H:He:C:N:O = 44:34:5:9:8, c.f.][]{abundposterusp}. For all these models log\,$g$\,=\,7 was assumed. 

\begin{figure}
	\includegraphics[width=\columnwidth]{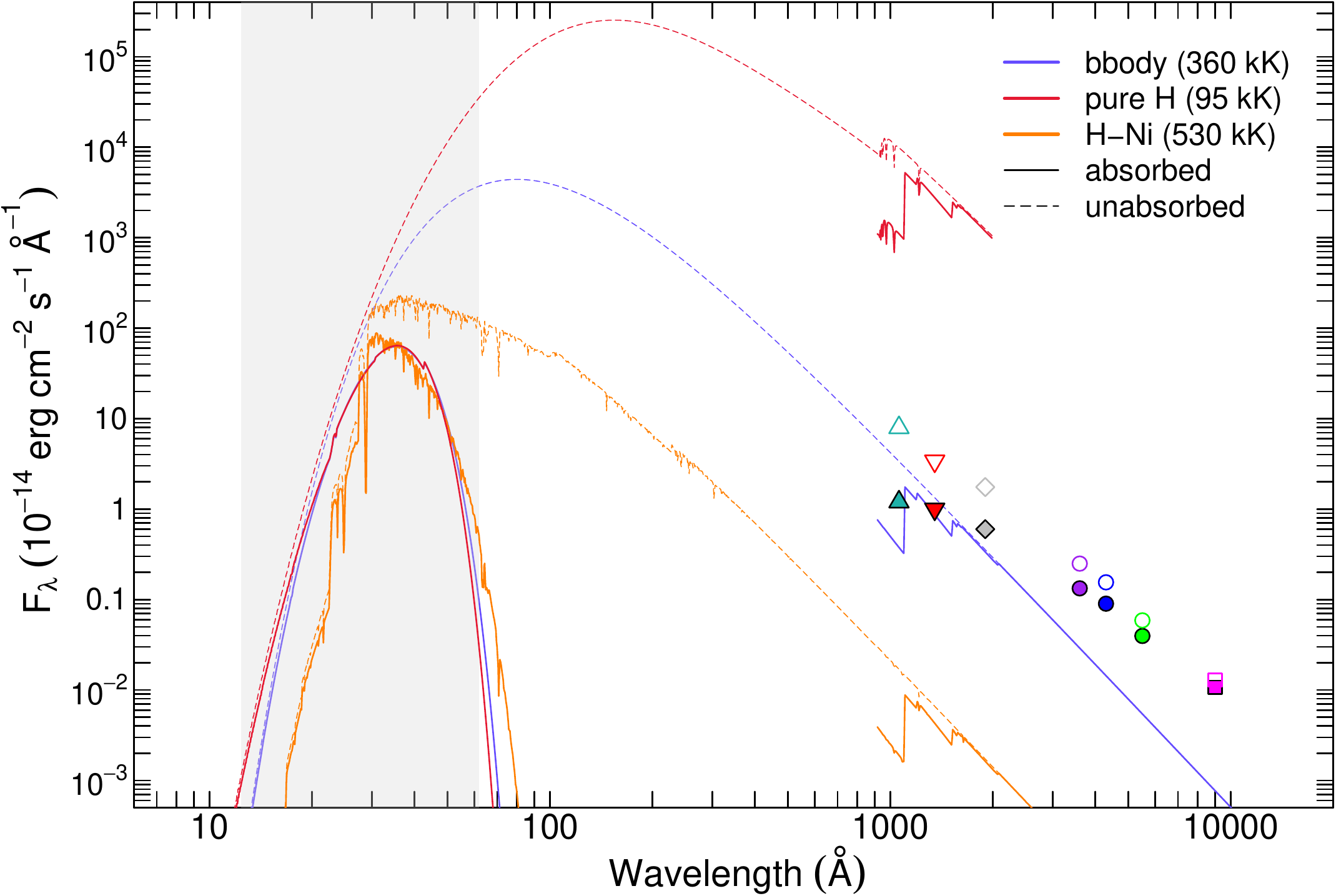}
    \caption{Model spectra from \texttt{bbody}, \texttt{pure\,H} and \texttt{H-Ni} best fits to data of ObsID 0506531501. Solid and dashed lines indicate absorbed and unabsorbed cases, respectively. Filled symbols are historical measurements for \Cal\ in different wavelengths. In increasing order of wavelength: sea green triangle (up) corresponds to data from FUSE \citep{2004AJ....127..469S}; red triangle (down) from HST \citep{1998A&A...333..163G}; grey diamond from IUE \citep{1987ApJ...321..745C}; coloured circles (UBV) from Curtis Schmidt telescope \citep{2002ApJS..141...81M} and magenta rectangle from OGLE-III \citep{2013MNRAS.432.2886R}. Unfilled symbols are these same measurements corrected for reddening with $A_{\text{V}}$\,=\,0.43\,mag. Energy range of analysis (0.2--1.0\,keV) is highlighted in grey.}
    \label{fig:05}
\end{figure}

The analysis was expanded by gathering from the literature measurements of \Cal\space at different wavelengths in order to evaluate how they relate to the fluxes derived from the X-ray spectral fitting. These measurements are presented in Figure\,\ref{fig:05}, along with the model spectra from the \texttt{bbody}, \texttt{pure\,H} and \texttt{H--Ni} fits to data of Obsid 0506531501. We opted to show these two atmosphere models as they represent two extreme cases regarding metallicity and best-fitting temperature. As one would notice, the models applied to the X-ray data were extrapolated to longer wavelengths; this was done by simply extending the energy range of the comparison (as far as each model grid allowed) \textit{after} the fit in X-rays is obtained, i.e. maintaining the best-fitting parameters. Absorbed (i.e. \NH\space with best-fitting values) and unabsorbed (i.e. \NH\space values set to zero) cases of each model are plotted in solid and dashed lines, respectively. The infrared/optical/UV data points (filled symbols) used to build this spectral energy distribution (SED) are referenced in the figure's caption. Also shown (unfilled symbols) are these fluxes corrected for optical/UV extinction ($A_{\text{V}}$), computed through the relation \NH/A$_{\text{V}}$\,=\,2.87\,$\times$\,10$^{21}$\,cm$^{-2}$\,mag$^{-1}$ \citep{2016ApJ...826...66F} and using the extinction law from \cite{1989ApJ...345..245C}. The estimated \NH\,from the \texttt{bbody} fit taking into account Galactic and LMC absorption, \NH\,=\,1.25\,$\times$\,10$^{21}$\,cm$^{-2}$, corresponds to $A_{\text{V}}$\,=\,0.43\,mag. We see that -- although consistent with each other in the X-ray analysis range (region highlighted in grey) -- the fluxes derived from the three models applied to the X-ray spectra differ greatly when extrapolated to longer wavelengths, being the flux computed from the phenomenological blackbody the most consonant with \Cal\space historical measurements. 

\begin{figure}
	\includegraphics[width=\columnwidth]{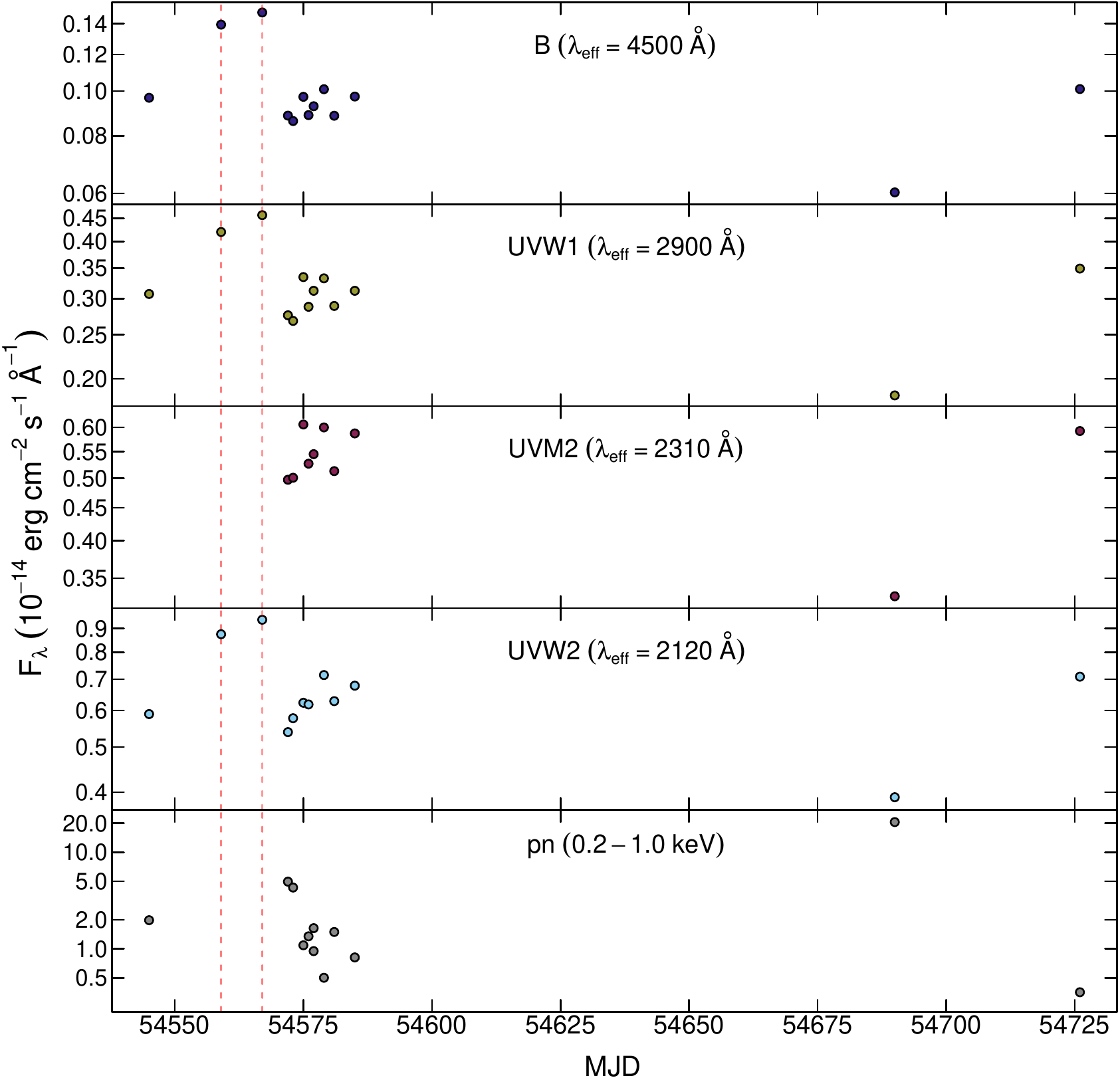}
    \caption{Simultaneous flux measurements for different OM filters (four first panels) and pn (bottom panel). OM data were taken from XMM-SUSS catalogues; X-ray fluxes were calculated from an absorbed blackbody fit to the spectra. 
    Vertical red dashed lines indicate the occasions when \Cal\space was not detectable in X-rays.}
    \label{fig:06}
\end{figure}

Aiming to probe on the correlation of the X-ray variability of \Cal\space regarding longer wavelengths, we browsed the \textit{XMM-Newton} Serendipitous Ultraviolet Source Survey \citep[XMM-SUSS, e.g.][]{2012MNRAS.426..903P} for the optical/UV flux measurements from the observations analysed in our study. The XMM-SUSS catalogues\footnote{\url{https://www.cosmos.esa.int/web/xmm-newton/om-catalogue}} comprise data from the \textit{XMM-Newton} Optical Monitor \citep[OM,][]{2001A&A...365L..36M}, a telescope that is co-aligned with the X-ray cameras and thus enables strictly simultaneous X-ray and optical/UV data examination. Figure\,\ref{fig:06} presents fluxes versus date for the 14 observations (ObsIDs 0506530201 to 0506531601, see Table\,\ref{tab:01}) in which OM measurements were available in a greater number of passband filters. First four panels show OM data for different filters (effective wavelengths are indicated) and the bottom panel shows the X-ray absorbed flux (0.2--1.0\,keV), computed after the \texttt{bbody} model fit to each pn spectrum. The two occasions when \Cal\space was not detectable in X-rays during the contemplated period are indicated (vertical red dashed lines through all panels); missing data points from OM panels (in regard to X-ray data points) simply means that the current filter was not used during the observation. These simultaneous measurements show that there is clearly an anti-correlation between the X-ray and UV/optical emission of the source in a timescale of days to weeks. It is worth pointing out that the amplitude in X-rays varies by a factor up to 20, while for longer wavelengths this factor is not much larger than 2. 

\begin{figure}
	\includegraphics[width=\columnwidth]{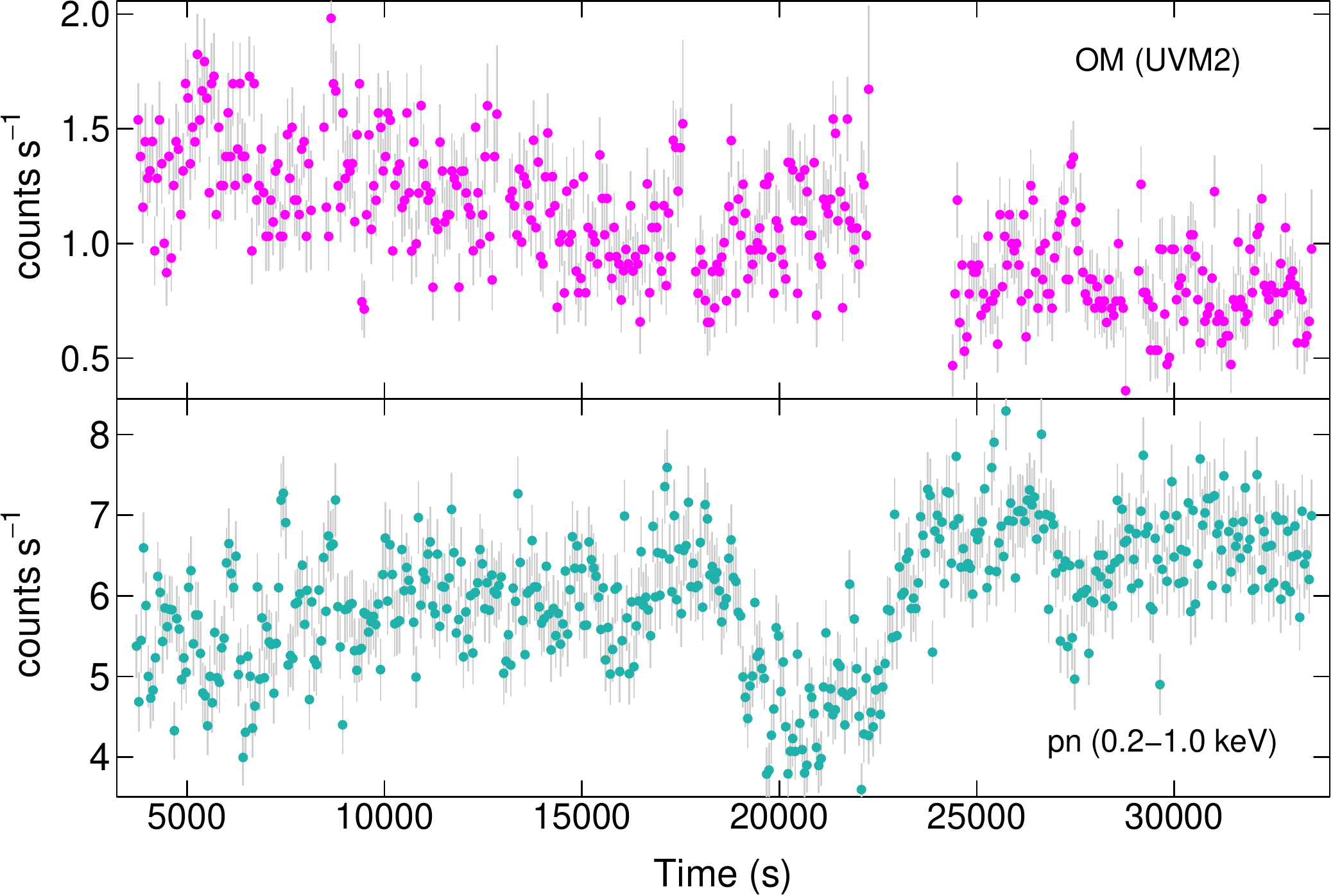}
    \caption{Light curves for ObsID 0506531701. \textit{Upper panel}: OM's UVM2 filter ($\lambda_{\text{eff}}$\,=\,2310\,$\r{A}$). \textit{Bottom panel}: EPIC-pn (0.2--1.0\,keV). The time binning for both is 60 seconds.}
    \label{fig:07}
\end{figure}

In an attempt to verify if such anti-correlated behaviour is perceptible on the timescale of a single observation or less, we turn to \mbox{ObsID~0506531701}, the lengthiest uninterrupted \textit{XMM-Newton} exposure of \Cal. 
We show in Figure\,\ref{fig:07} the simultaneous net count rate for this observation from OM's UVM2 filter ($\lambda_{\text{eff}}$\,=\,2310\,$\r{A}$, upper panel) and EPIC-pn (0.2--1.0\,keV, bottom panel). As no rigorous timing analysis was intended for the OM data, we used the background-subtracted light curves that are readily available for download at the \textit{XMM-Newton} Science Archive. There were eight continuous exposures taken in OM's fast mode with filter UVM2, each lasting for approximately 4.4\,ks, with gaps (due to overheads) of about 300 seconds between them. The longer gap, of $\sim$\,2000 seconds (seen at Time\,$\approx$\,22,500\,s), is likely due to a ground station handover\footnote{\url{https://xmm-tools.cosmos.esa.int/external/xmm_user_support/documentation/uhb/ommodes.html}}. The X-ray light curve was extracted as explained in Section\,\ref{sect:02}. Data points from pn were clipped to match the start and end time of that available from OM exposures; both light curves are shown with time bins of 60 seconds. There are three interesting features to pay attention to in these simultaneous light curves: a slight ascending trend in the X-ray counts  (until Time\,$\approx$\,17\,ks) that occurs concomitantly with a slight descending trend in the UV counts; a few excursions, somewhat similar to flickering events (e.g. at Time\,$\approx$\,5.5, 21, 27\,ks), which are present in both curves and seem to be anti-correlated; and a rise in UV that occurs almost simultaneously with the beginning of the X-ray decline towards the dip (at Time\,$\approx$\,18\,ks). Unfortunately,  the further increase in pn counts and decrease in UV coincides with a gap in OM data due to the ground station handover. 

At last, we computed the Lomb-Scargle periodograms for all 18 observations's light curves (0.2--1.0\,keV, with the 5\,s binning) and found that only in 5 of them the prominent peak previously reported around 65--72\,s surpassed the 90\% true alarm probability \citep[e.g.][]{1982ApJ...263..835S}. They are ObsIDs 0500860301, 0500860501, 0500860601, 0506531501 and 0506531701, with modulations at 67.5\,$\pm$\,0.3, 72.7\,$\pm$\,0.2, 67.7\,$\pm$\,0.1, 66.9\,$\pm$\,0.3 and 65.17\,$\pm$\,0.04 seconds, respectively. These values are in agreement with those reported in  \cite{2014MNRAS.437.2948O} and \cite{2017MNRAS.467.2797O} for the same observations. It is worth mentioning that virtually the same periods and significance values were found when the search was performed for light curves extracted in the 0.2--0.4\,keV range. Conversely, the signal could not be detected for the complementary (0.4--1.0\,keV) band, which may be simply associated with the poor photon statistics in this particular range and not necessarily with an energy-dependent modulation. For instance, when comparing the phase-folded light curves for the 0.2--0.4 and 0.4--1.0\,keV bands, we find that the modulation amplitudes are consistent to a factor of $\sim$\,2, although with shapes less defined in the latter case. Moreover, with better signal-to-noise ratio light curves from \textit{NICER}, \cite{2022ApJ...932...45O} reported similar amplitude modulations of the $\sim$\,67\,s signal in both bands.

\section{Discussion}
\label{sect:04}

\subsection{Modelling the X-ray low resolution spectra of \Cal}

A rigorous fit to \textit{Chandra} and \textit{XMM-Newton} high resolution spectra (65--20\,$\AA$; $\sim$\,0.2--0.6\,keV) of \Cal\space was performed by \cite{2005ApJ...619..517L} with a sophisticated NLTE atmosphere model specifically designed for the object. A fine match between their model (which is not public) and the emission and absorption features of \Cal, that are discernible in the grating data, was obtained with best-fitting parameters of T\,=\,550\,kK and log\,$g$\,=\,8.5. Due to the low energy resolution of EPIC-pn camera, compared to the grating instruments, such absorption and emission features are unresolved in the spectra we analysed, and the X-ray continuum -- i.e. the spectrum overall shape -- of \Cal\space is found to be well described by a blackbody. A reasonable fit could be achieved by a phenomenological blackbody model for 17 out of 18 spectra; the exception was the one extracted from the observation whose length is almost half the orbital period (see, e.g., Figure\,\ref{fig:07}), a span long enough that, given the source's variability, produces a composite spectrum likely not well defined by a single temperature. The effective temperatures provided ($\sim$\,30\,eV$\approx$\,350\,kK) are consistent with those reported by \cite{2013MNRAS.432.2886R}, for the same data, and lie within the broad range of values previously reported for \Cal\space($\sim$\,20--50\,eV).

A pure hydrogen atmosphere model, the only one amongst the publicly available NLTE models that was able to fit the data, demands much lower temperatures ($\sim$\,8\,eV$\approx$\,95\,kK) to describe the level of the spectra. The application of this model can be seen as exploratory, given that \cite{1987ApJ...321..745C} had already reported a hydrogen-poor composition from optical/UV spectroscopy of \Cal. Moreover, with the temperatures and unabsorbed fluxes obtained from our fits to the X-ray spectra (Figure\,\ref{fig:02}), and assuming a distance of 50\,kpc to the LMC \citep[e.g.][]{2019Natur.567..200P}, we would find radii of roughly 0.01 to 0.05 R$_{\odot}$ for the \texttt{bbody} and up to 1 to 2 R$_{\odot}$ for the \texttt{pure\,H} model. The value from the blackbody already surpasses the radius of low-mass WDs (although an inflated photosphere scenario could be considered), but the value derived from the pure hydrogen atmosphere model is just completely unrealistic. 

On the other hand, none of the three publicly available NLTE grid models that include heavier elements, \texttt{He+CNO}, \texttt{H--Ca} and \texttt{H--Ni}, could provide a reasonable fit to the \Cal\space X-ray data we analysed. 
The \texttt{He+CNO} grid has a maximum limit temperature of 190\,kK. When a fit is attempted those hottest models are still nowhere near describing the spectrum.
The other two models (\texttt{H--Ca} and \texttt{H--Ni}), whose grids comprise temperatures up to 1000\,kK, had better performances but still failed to properly describe the data, particularly beyond 0.45--0.5\,keV. 
For the \texttt{H--Ca} model, one may choose between two abundance ratios: solar and that of the Galactic halo. In both cases the spectrum of \Cal\space for energies above $\sim$\,0.5\,keV is underestimated. The halo abundances, that provided a better -- but still far from good -- fit was displayed in Figure\,\ref{fig:03}. The \texttt{H--Ni} model contemplates additional elements and is available for a few abundance ratios, but was intended for a specific object other than \Cal. The best possible fit for this grid was selected to be presented in Figure\,\ref{fig:03}. The inconsistency between this model and data begins at an even lower energy ($\sim$\,0.45\,keV) and also grows much larger, reaching a data to model ratio of a factor 100. Similar issue in fitting the more energetic part of \Cal\space X-ray low resolution spectra with atmosphere models has been previously reported. One of the first attempts was carried out by \cite{1998A&A...332..199P} with data (0.1--0.8\,keV) from \textit{BeppoSAX}. The applied NLTE model provided acceptable fits for both solar and 0.25 solar metallicities. The authors found, however, that the fit quality was actually worse than a blackbody, mainly because the atmosphere model underestimated the spectrum at energies $\gtrsim$\,0.4\,keV. The effective temperatures obtained were about 350--400\,kK for a few values of gravity between log\,$g$\,=\,8--9.  More recently, \cite{2022ApJ...932...45O} attempted to fit \textit{NICER} spectra (0.2--1\,keV) of \Cal\space with the \texttt{H-Ca} model. The authors, that opted for the halo abundance ratio, found that the model underpredicted the source's emission above 0.5\,keV. Best-fitting temperature was $\sim$\,500\,kK for log\,$g$\,=\,9.  

As illustrated in Figure\,\ref{fig:04}, the choice of abundance ratio highly modifies the output continuum and hence the uniqueness of a particular SSS certainly demands that models distinctively targeted to describe it are built \citep[as done, e.g., by][]{2005ApJ...619..517L}. For that, accurate composition knowledge would be very useful and spectroscopic observations at longer wavelengths (e.g. optical, infrared), something that has not happened in a while for \Cal, should be encouraged.  

\subsection{The spectral energy distribution}

Although the historical flux measurements displayed in Figure\,\ref{fig:05} were taken in different epochs and \Cal\space is known to vary in these wavelengths, such variability hardly exceeded a factor 2--2.5 (see references for each, and also Figure\,\ref{fig:06}), which is just slightly larger than the size of the symbols that represent such measurements in the plot. As for the X-ray energy distributions plotted, they come from fitting the \textit{XMM-Newton}/pn observation which had the highest flux, meaning that for none of the other observations analysed the \texttt{bbody} model best-fit extrapolation exceeded the observed flux in longer wavelengths. In the same way, the flux extrapolations shown for the \texttt{pure\,H} and the \texttt{H--Ni} models are also the ceiling levels in our sampling. The \NH\space values certainly play an important role when estimating unabsorbed fluxes, and different levels in longer wavelengths could be achieved by varying this parameter from the best-fitting values. We remind, none the less, that the extrapolations were computed from each model's output parameters that best fitted the X-ray data (0.2--1.0\,keV). That said, although able to approximately match the observed flux in the X-ray energy range, the spectral energy distribution of the enhanced metal (\texttt{H--Ni}) and the pure hydrogen atmosphere (\texttt{pure\,H}) models largely underestimate and overestimate, respectively, \Cal\space measurements in longer wavelengths. 

By looking at the SED of less absorbed SSS on multi-wavelength scales one may discriminate between models with good agreement to X-ray data alone. On the other hand, the low energy SED is prone to contributions from different sources in the binary system. The simple 360\,kK blackbody model for the soft X-ray emission presented in Figure\,\ref{fig:05} roughly describes the SED level observed in the UV, with an increasing deficit towards the optical and near IR. Such an observed flux above the blackbody model may be understood as the central source actual emission from a hot atmosphere and/or the power-law contribution from an accretion disc -- a larger source with lower temperatures and log\,$g$.

\subsection{X-ray and optical/UV anti-correlated variability}

We are aware that the values provided by a simple blackbody model fit may not represent realistic physical parameters, but the variation between them, seen from several different epochs, is certainly useful in obtaining valuable information about the evolution of the source. For instance, we find that there is a mild positive correlation between the X-ray fluxes observed from \Cal\space and the effective temperatures derived from spectral analysis (e.g. Pearson's coefficient $\sim$\,0.6). That, together with the known fact that X-ray and longer wavelengths' emissions are anti-correlated in \Cal, hints that the system's X-ray source cools as other emitting components of the system become brighter. This had already been shown and discussed by \cite{2013MNRAS.432.2886R} for the same set of X-ray data we analyse here and almost-concurrent infrared measurements of \Cal\space from OGLE-III.

To our knowledge, Figure\,\ref{fig:06} presents the most simultaneous optical, UV and X-ray flux measurements ever reported for \Cal. It makes it explicit that the emission in X-rays is, observed in a time scale of days, straight away anti-correlated with that in both optical and UV.

\begin{figure}
	\includegraphics[width=\columnwidth]{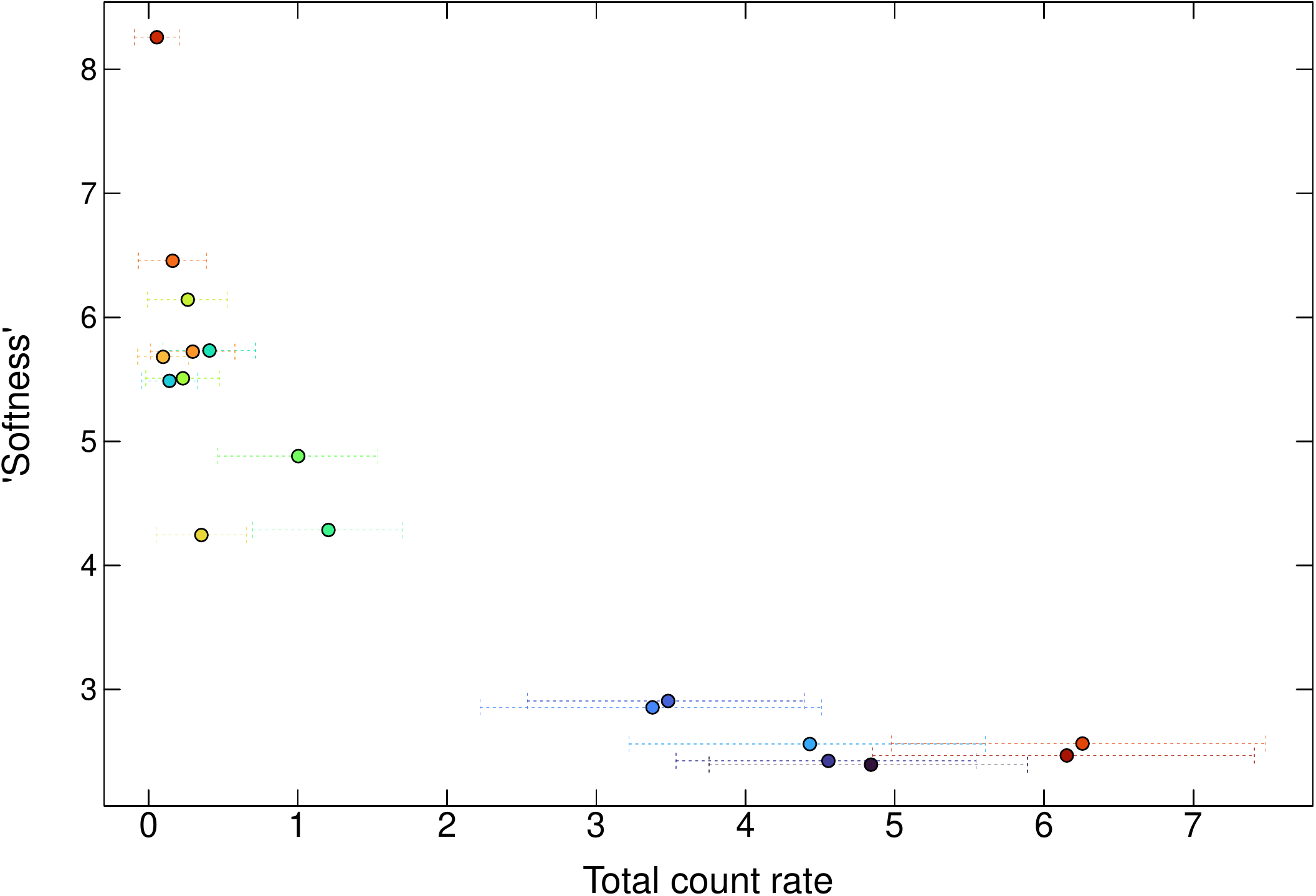}
    \caption{'Softness', defined as the ratio of the mean count rate in bands 0.2--0.35 and 0.35--1.0\,keV, versus the mean total count rate in the whole band (0.2--1.0\,keV). Values were calculated from each observation's light curve extracted in the respective energy bands. Horizontal bars are the count rate standard deviation. Colour codification is as in Figure\,\ref{fig:01}.}
    \label{fig:08}
\end{figure}

At a first glance, the amplitude of the X-ray and optical/UV anti-correlated variations may suggest an origin related to absorption by gas within the primary Roche lobe or in a circumbinary ejecta, with the amplitude in X-rays being much larger (by almost a factor of 10) than in the UV or optical. However, the hardness ratio in soft X-rays does not corroborate such a hypothesis, as first pointed out by \cite{2022ApJ...932...45O}. They reported, from analysing 6 \textit{NICER} observations of \Cal, that the  emission becomes softer as the source is fainter (see their Table~2). We adopted their "softness ratio" definition and calculated for each of the observations analysed here the ratio of the mean count rates (from extracted light curves) in a softer (0.2--0.35\,keV) and in a less soft band (0.35--1.0\,keV) to build Figure\,\ref{fig:08}. The numerous 'softness' measurements shown, taken at several epochs, are in fact not consistent with photoelectric absorption and would thus require an explanation consistent with the observed X-rays--optical/UV anti-correlation. 

Different scenarios to explain the optical pattern variability (and, at some point, the corresponding X-ray anti-correlated emission) have been proposed, most based on the WD's photosphere expansion/contraction model \citep{1996ApJ...470.1065S}, being the 'trigger' for the photospheric radius adjustments -- e.g. changes in the mass accretion rate or in the burning rate \citep{2002A&A...387..944G} --  one of the main questions. Other scenarios, such as accretion disk instability or thermonuclear runaways without mass ejection \citep[e.g.][]{2005ApJ...623..398Y}, that have provided appropriate description of the behaviour of some SSS \citep[e.g.][]{2019ApJ...879L...5H}, are also possibilities. It should be noted, though, that no expressive outburst of \Cal\space has been observed; long-term light curves in optical and infrared \citep[e.g. Figure 1 of][]{2013MNRAS.432.2886R} show that the emission in these wavelengths is certainly variable, in a time scale of weeks to months -- however, with an amplitude that does not exceed $\sim$\,1.5\,mag. Our analysis does not allow to explore these scenarios and thus draw conclusions on the origin of the high and low optical states emission observed. Observations in various multiwavelengths along a state transition, in tandem with an accretion disk tomography analysis, may help elucidate these hyphoteses.

On much shorter time-scales, such as those presented in Figure\,\ref{fig:07}, one may see anti-correlated flickering events in simultaneous X-ray and UV data, suggesting that this effect occurs over a broad time-frequency range, although not necessarily sharing a common origin. 

\section{Conclusions}
\label{sect:05}

We have retrieved public available data from the \textit{XMM-Newton} satellite to revisit and discuss on relevant aspects of the supersoft X-ray source \Cal. After carefully and systematically reducing the X-ray data by performing a series of tasks to guarantee that the output spectra were reliable for analysis, blackbody and publicly available atmosphere models were applied in an attempt to describe the object's soft X-ray spectral shape. Historical flux measurements of \Cal\space in other wavelengths were used to build an extended spectral energy distribution, allowing a straightforward comparison of such measurements with the fluxes derived from the X-ray modelling. Additionally, optical/UV measurements from the \textit{XMM-Newton} itself -- and thus simultaneous with the X-ray data -- were used to assess the source's anti-correlated variability in X-rays with respect to these other wavelengths. The main findings obtained from these three approaches are summarised below. 

\begin{enumerate}

\item The X-ray spectral shape of \Cal\space is well described by a blackbody model with temperatures of about 320--370\,kK. A pure hydrogen atmosphere model also fits the data, although providing much lower effective temperatures (80-105\,kK). Other public available models, that contemplate heavier elements, demand higher temperatures ($\gtrsim$\,420\,kK) and do not properly reproduce the spectra we analysed. The main reason, we assert, is that as these pre-calculated grids were intended for other sources, they carry specific abundance ratios and ranges of log\,$g$ and effective temperatures that may not precisely represent the parameters of \Cal.

\item Although not to be taken as the nominal temperature value of the source, due to the simplicity and phenomenology of the model, there is a very good agreement between the spectral energy distribution of a $\sim$\,360\,kK blackbody and measurements of \Cal\space previously reported in UV, optical and near infrared.

\item \Cal\space exhibits, in a time scale of days to weeks, an anti-correlated behaviour in X-rays with respect to optical/UV, the former with a much larger variation amplitude. When analysing the hardness ratios of each observation one can see that the X-ray emission becomes harder as the source becomes fainter. Such finding contradicts a scenario of variable absorption. The source also shows a X-ray--UV anti-correlated tendency/variations in a time scale of minutes.

\end{enumerate}

\section*{Acknowledgements}

PES acknowledges PCI/INPE/CNPq for financial support under grant \#300320/2022-1. MPD thanks support from CNPq under grant \#305033. The authors thank an anonymous referee for her/his helpful comments.

\section*{Data Availability}

The data underlying this article will be shared on reasonable request to the corresponding author.
 











\bsp	
\label{lastpage}
\end{document}